\documentclass[english]{article}

\usepackage[T1]{fontenc}
\usepackage[utf8]{inputenc}
\usepackage{babel}
\usepackage{csquotes}
\usepackage{hyperref}
\usepackage{ragged2e}
\usepackage[style=ieee, citestyle=numeric-comp, backend=biber]{biblatex}
\usepackage{graphicx}
\usepackage{float}
\usepackage{nameref}
\usepackage[top=1cm,bottom=1.5cm,left=3cm,right=3cm,includeheadfoot]{geometry}
\usepackage{caption}
\newcommand{\beginsupplement}{%
        \setcounter{table}{0}
        \renewcommand{\thetable}{S\arabic{table}}%
        \setcounter{figure}{0}
        \renewcommand{\thefigure}{S\arabic{figure}}%
     }

\DeclareUnicodeCharacter{0301}{\'{e}}

\begin{document}


\title{Tomographic Volumetric Additive Manufacturing of Silicon Oxycarbide Ceramics}

\maketitle


\author{Max Kollep, }
\author{Georgia Konstantinou, }
\author{Jorge Madrid-Wolff, }
\author{Antoine Boniface, }
\author{Pradeep Vallachira Warriam Sasikumar, }
\author{Gurdial Blugan, }
\author{Paul Delrot, }
\author{Damien Loterie, }
\author{Christophe Moser*}



M. Kollep, G. Konstantinou, J. Madrid-Wolff, Dr. A. Boniface, Prof. C. Moser \\
Laboratory of Applied Photonics Devices,
Ecole Polytechnique Federale Lausanne (EPFL),
CH-1015 Lausanne, Switzerland \\

Dr. P. V. W. Sasikumar, Dr. G. Blugan\\
Laboratory for High Performance Ceramics,
Swiss Federal Laboratories for Material Science and Technology (Empa), 
CH-8600 Duebendorf, Switzerland \\

Dr. P. Delrot, Dr. D. Loterie \\
Readily3D SA,
EPFL Innovation Park Bâtiment A,
CH–1015 Lausanne, Switzerland \\

$*$ Email Address: christophe.moser@epﬂ.ch

\textit{keywords - } Ceramics, volumetric additive manufacturing, polymer derived ceramics, 3D printing, SiOC, preceramic polymers

\begin{abstract}
    Ceramics are highly technical materials with properties of interest for multiple industries. Precisely because of their high chemical, thermal, and mechanical resistance, ceramics are difficult to mold into complex shapes. A possibility to make convoluted ceramic parts is to use preceramic polymers (PCP) in liquid form. The PCP resin is first solidified in a desired geometry and then transformed into ceramic compounds through a pyrolysis step that preserves the shape. Light-based additive manufacturing (AM) is a promising route to achieve solidification of the PCP resin. Different approaches, such as stereolithography, have already been proposed but they all rely on a layer-by-layer printing process which sets limitations on the printing speed and object geometry.
Here, we report on the fabrication of complex 3D centimeter-scale ceramic parts by using tomographic volumetric printing which is fast, high resolution and offers a lot of freedom in terms of geometrical design compared to state-of-the-art AM techniques. First, we formulated a photosensitive preceramic resin that was solidified by projecting light patterns from multiple angles. Then, the obtained 3D printed parts were converted into ceramics by pyrolyzing them in a furnace. We demonstrate the strength of this approach through the fabrication of dense microcomponents exhibiting overhangs and hollow geometries without the need of supporting structures, and characterize their resistance to high heat and harsh chemical treatments. 

\end{abstract}

\section{Introduction}

\justifying

\paragraph{} Over the past decades, ceramics have attracted much interest for their superior properties including hardness, durability and stability in extreme environments. They meet fabrication needs in various fields ranging from transportation industry (e.g. diesel engines) to the energy sector (e.g. nuclear) but also environment, defense, aerospace and in the medical sector (e.g. ceramic thermal barrier coatings, filters, lightweight space mirrors, hip or knee implants) \cite{greil2002advanced,JANSSEN20081369,He2018,Fu2019,deFaoite2011,MAKOWSKA2020134}.
However, the fabrication of complex ceramic parts remains very challenging. Because of their brittleness, conventional manufacturing processes, such as molding, are limited to simple object geometries as well as being costly and time-consuming \cite{chen20193d}. Additive manufacturing (AM) represents an attractive alternative. AM technologies offer more flexibility in terms of architecture and significantly reduce material waste. Also, they allow cost-effective production in a shorter time. 
In all the slurry-based additive manufacturing (AM) technologies being used for the fabrication of ceramics, the process starts with a liquid preceramic polymer (PCP) which is shaped in 3D. The result is a so-called green body, which undergoes a post-processing step of pyrolysis to result in the final ceramic  \cite{zocca2015additive,chen20193d,chen2021dense,chartier2013handbook,somiya2013handbook,JANSSEN20081369,he2021progress,liu20203d,Xu2019,MARTINEZCRESPIERA2011913,LU2016411,pelza2021additive}.

\paragraph{} PCPs have been used for the fabrication of polymer derived ceramics (PDCs) since the 1960s. Initially, resins sensitive to heat or light were used for the fabrication of disc-shaped green bodies, \cite{Xu2019,MARTINEZCRESPIERA2011913,LU2016411,SASIKUMAR201820961,Bernal2019} followed by micro-molding and injection molding \cite{GROSSENBACHER2015623,ZHANG1995729}. Through photopolymerization, preceramic polymers can also be solidified into a rigid green body by stereolithography (SLA). It enables the additive manufacturing of more complex shapes, that are preserved after pyrolysis through the polymer-to-ceramic transformation \cite{eckel2016additive}. Lithoz GmbH fabricated SiOC ceramic parts by using light based DLP printing. The latter uses a Digital Micromirror Device to project 2D patterns inside a transparent vat with the viscous photocurable ceramic suspension, applied as a thin leveled film by a combination of vat rotation and wiper blade \cite{schwentenwein2014lithography}.

\begin{figure}[t!]
    \centering
    \includegraphics[width=1\textwidth]{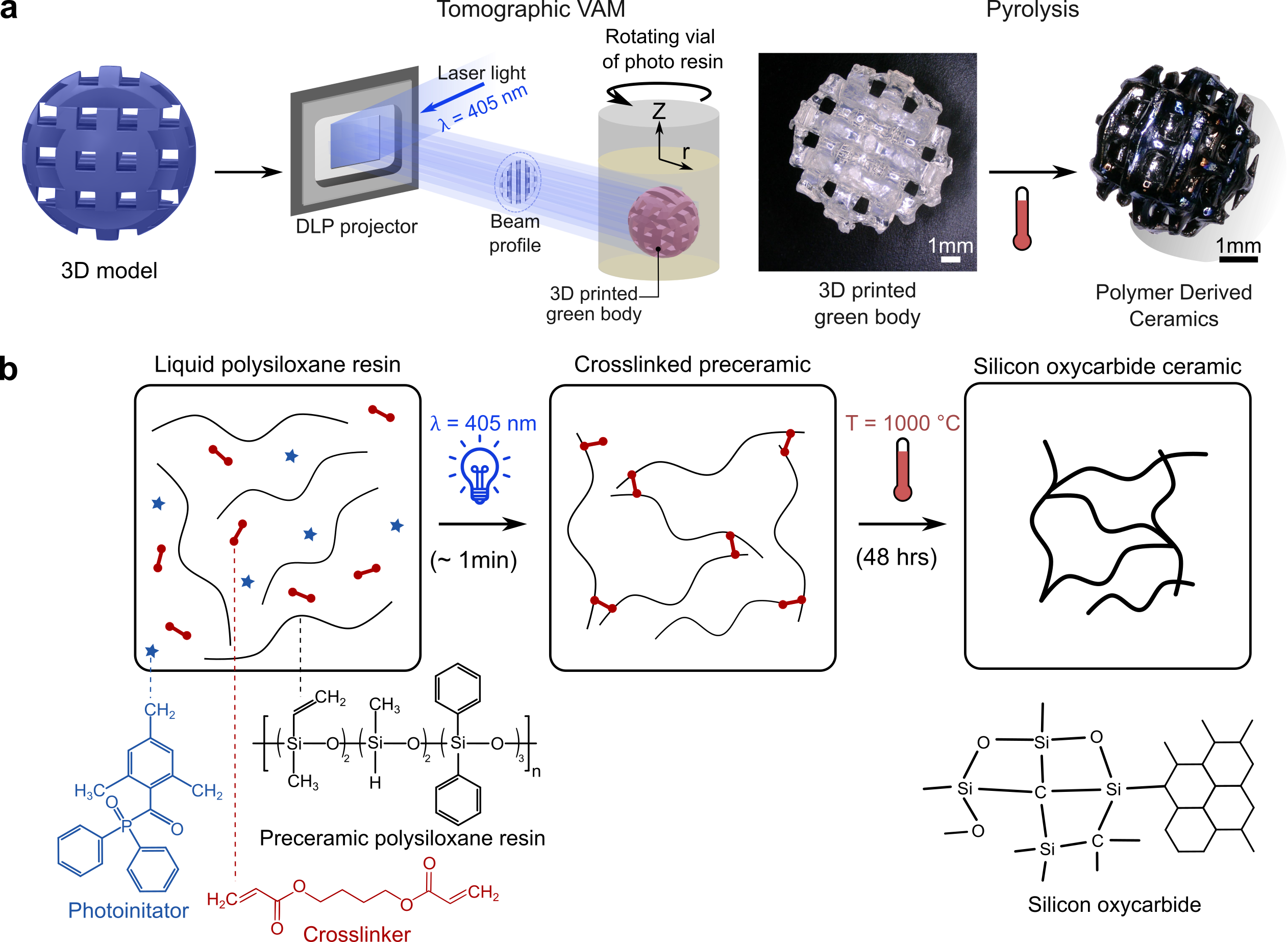}
    \caption{\textbf{a}. In \textbf{tomographic volumetric additive manufacturing of silicon oxycarbide ceramics}, the 3D model of the desired part is used to calculate a set of light patterns which are projected onto a rotating vial filled with a photo-curable preceramic resin. The resulting solid green body is retrieved from the liquid resin, and pyrolized at $\textit{T} = 1000 ^{\circ} C$. \textbf{b}. Schematic representation at the molecular level. A polysiloxane preceramic resin is mixed with a crosslinker and a photoinitiator. After excitation with blue light, a stiff network of polymerized and crosslinked polysiloxane chains forms the green body. A 48h pyrolysis cycle burns out most of the organic components and transforms the part into a silicon oxycarbide ceramic amorphous network \cite{mera2013polymer}.}
    \label{fig1}
\end{figure}

Based on a similar technology but with a blended formulation of two polysiloxanes, Schmidt et al. were able to print woodpiles, 2D and 3D grids, cork screws, honeycombs, micro-lattices or Kelvin cell structures which all maintain their initial shape during pyrolysis at temperatures of 1000$^{\circ}$C  \cite{schmidt2018digital,Colombo2010}. PDCs with micrometer resolution were first demonstrated  using two-photon lithography (2PL) by \cite{park2009fabrication} and later more complex structures were also reported with the Nanoscribe 3D printer using higher pyrolysis temperatures \cite{brigo20183d}. A low-shrinkage preceramic resin ($\sim$ 30 \%) was also cured using the Nanoscribe system and resulted in fully dense and crack-free 2PL-PDCs in \cite{KONSTANTINOU2020101343}. Multiscale ceramic parts were demonstrated by combining DLP and 2PL, for which micrometer resolution was obtained via the latter on the centimeter scale object fabricated by the former \cite{schmidt2019multiscale}. Still, overhangs and inner voids are challenging to produce using most AM technologies including vat SLA and DLP. To overcome this issue, the standard solution is to add supports struts during the printing process. Fine and manual post-processing must be then performed to remove those support structures, which limits the freedom of shape \cite{Jiang2018}.


\paragraph{}Volumetric 3D printing is a novel light-based technology that eliminates the need for support struts by printing the whole object at once within a vial of resin (see illustration in Figure \ref{fig1}a). Volumetric 3D printed objects are self-supported within the build volume and are built in a few tens of seconds as opposed to several minutes for DLP or SLA. The reason for this dereased building time is that the resin does not have to flow quickly to fill in the surface of the build plate when a new layer is cured. The essence of volumetric 3D printing is to produce a 3D light dose distribution within the volume of the photosensitive material using tomographic back-projections \cite{Kelly2019, Bernal2019, Loterie2020} or orthogonal dual-wavelength photopolymerization \cite{regehly2020xolography}. Tomographic volumetric additive manufacturing is not only faster but also produces isotropic homogeneous polymerized bodies because the whole object is polymerized at once rather than layer-by-layer \cite{Bernal2019}. Furthermore, this printing approach allows the fabrication of convoluted hollow structures and geometries with large overhangs which are unprintable with other AM techniques.
Recent progress in volumetric additive manufacturing now allow printing different materials including acrylic \cite{Loterie2020}, thiol-ene photoresins \cite{cook2020ThiolEne} or even scattering resins \cite{madrid2021scattering}, but the 3D printing of ceramics with a tomographic approach has never been reported to our knowledge.

\section{Results}

\justifying

\subsection{Preceramic resin}

\paragraph{} Here, we report on the volumetric additive manufacturing of SiOC ceramic components using a polysiloxane ceramic precursor with a crosslinker and a tomographic back-projection approach for photopolymerization. The resin used in the printer is composed of a polysiloxane back-bone (SPR 684) with 1,4-Butandiol-diacrylate (BDDA) as crosslinker. The photoinitiator Diphenyl-(2,4,6-trimethylbenzoyl)-phosphinoxide (TPO) is added as the light sensitive component. We found that the resin is highly transparent in the visible range, with most of its absorbance coming from TPO, as shown in Figure \ref{resin}.a-c. The acrylate-mediated photo-polymerization exhibits a thresholded non-linear response to light dose, which is fundamental in tomographic volumetric additive manufacturing (Figure \ref{resin}.b). Light attenuation can hinder the printability of cm-scale shapes (Figure \ref{resin}.d); thus we correct for the optical attenuation following the method described in \cite{madrid2021scattering}.

\paragraph{}The radical polymerization mechanism illustrated in Figure \ref{fig1}b begins with single-photon absorption by the photoinitiator (TPO). This generates the primary radicals (C-centered acyl and P-centered phosphinoyl radicals) after the $\alpha$-cleavage of the C-P bond \cite{lalevee2015dyes, eibel2018choosing}. The efficiency of the crosslinking at the propagation step is enhanced thanks to BDDA. In fact, the primary radicals of the initiation step (TPO) activate the radical polymerization of the BDDA by cleaving the methylene bond. The high reactivity of BDDA correspondingly assists the chain growth of the PCP by a similar mechanism of methylene cleavage. In this way, the crosslinking propagates to a direction perpendicular to the chain of the PCP. The termination step is ensured when the irradiation stops. After the pyrolysis step, amorphous SiOC is formed and the corresponding bonds are confirmed by X-ray photoelectron spectroscopy \cite{slavin2012cobalt, chastain1992handbook}.

\paragraph{}In the volumetric printer, parts are printed within rotating glass vials filled with the photo-curable resin as a set of light patterns are exposed onto it (\ref{fig1}a). The resin we present has a viscosity of \(873 mPas\), as shown in Figure \ref{viscosity}, which is high enough to prevent sinking of the polymerized part within the printing times of 30 to 60 seconds.

\paragraph{}The resin shows a nonlinear response to the light dose \cite{ligon2014OxygenInhibition}. This is caused by two effects: the gelation threshold of the resin \cite{flory1941molecular} and the presence of an inhibitor in the resin. The inhibitor reacts with the excited photoinitiator, preventing the initiation of the polymerization process. This inhibitor shifts the polymerization to higher light doses since it must be locally depleted in order for the polymerization to start. In our case, the inhibition is caused by the oxygen naturally dissolved from the atmosphere in the resin. This nonlinear response is crucial for the volumetric printing process because it creates a threshold of light dose that has to be surpassed to start the polymerization. Thanks to it, the light projected from the DMD can penetrate the resin without polymerizing it directly at the edges of the vial. Only in the center of the vial where the light dose is cumulated from multiple exposures at different angles does it surpass the threshold and polymerizes the resin.

\subsection{Geometrical characterization of 3D printed ceramic parts}
\begin{figure}[b!]
    \centering
    \includegraphics[width=130 mm]{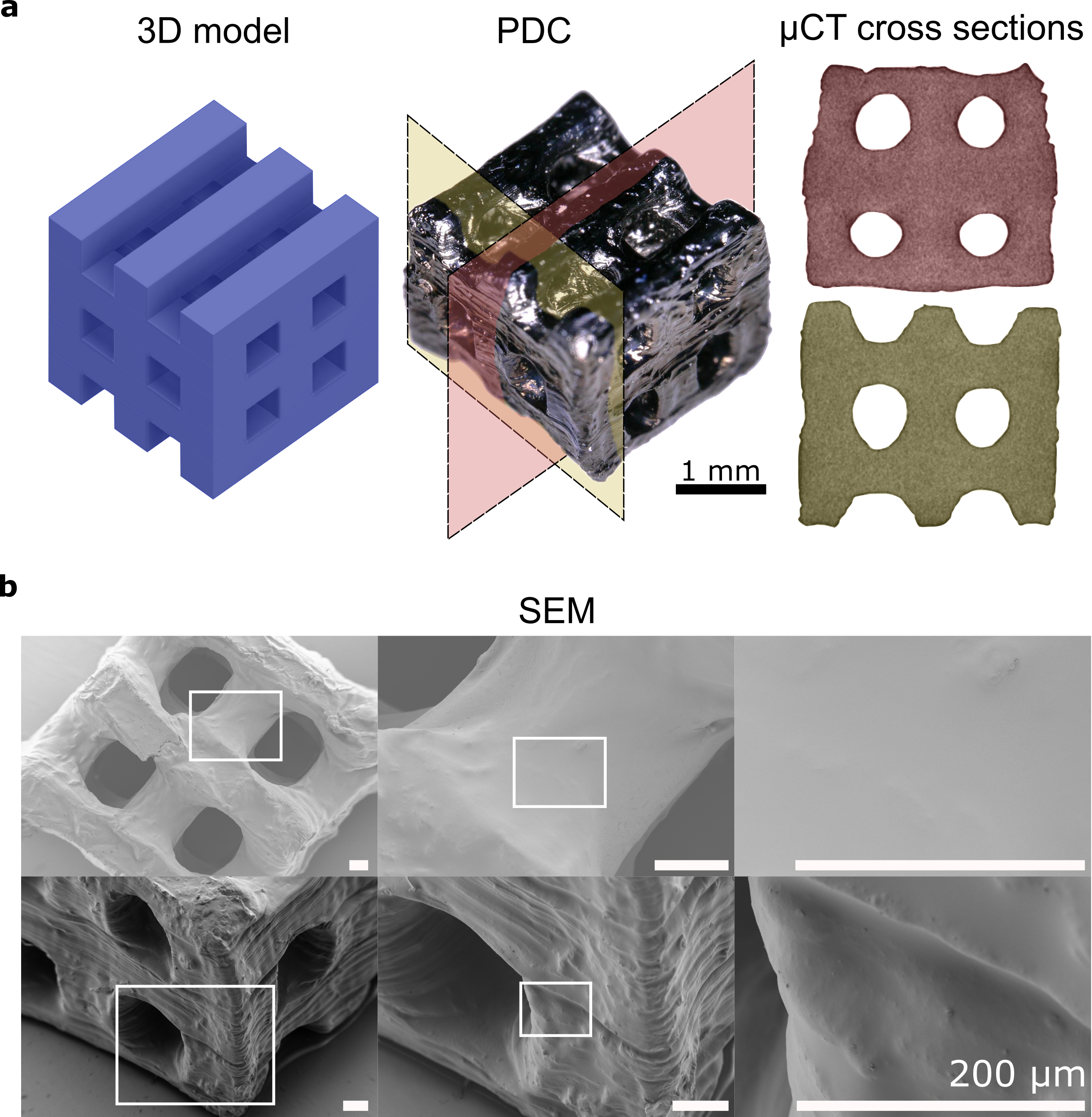}
    \caption{\textbf{Density and smoothness. a.} 3D model, microscope image and cross sections from micro Computed Tomography images of a 5-level woodpile. Scalebar 1 mm. \textbf{b.} Scanning Electron Microscopy images of the printed parts and their surfaces. Scalebars  200 µm. }
    \label{fig2}
\end{figure}

\paragraph{}To first illustrate the volumetric method, we fabricated polymer-derived ceramic woodpiles measuring 5 mm $\times$ 5 mm $\times$ 5 mm with 5 layers of rods of square cross section measuring 1 mm $\times$ 1 mm and spaced out by voids of the same dimensions, as shown in Figure \ref{fig2} a and b. Micro-CT scans of the prints show that they are fully dense with no signs of cracks or porosity throughout the bulk of the PDC (supporting videos 1 and 2). The parts also retain the design voids inside, even after pyrolysis. Figure \ref{fig2}b shows a series of scanning electron microscopy images of the woodpile structure, highlighting the surface quality of the prints. The fabricated ceramics are smooth and crack-free. It is visible that the vertical walls of the PDC exhibit some striations when compared to the horizontal ones. These striations are typical of volumetric additive manufacturing and might come from self-writing waveguide effects \cite{shoji1999optically, rackson2021StriationFreeVAM}. We observe that the surface smoothness is higher compared to ceramic structures occurring from Digital Light Processing \cite{liu20203d}, \cite{he2020polymer}. Additionally, volumetric additive manufacturing simplifies post-processing of the green bodies due to the absence of a build plate or support structures.

\begin{figure}[b!]
    \centering
    \includegraphics[width=150 mm]{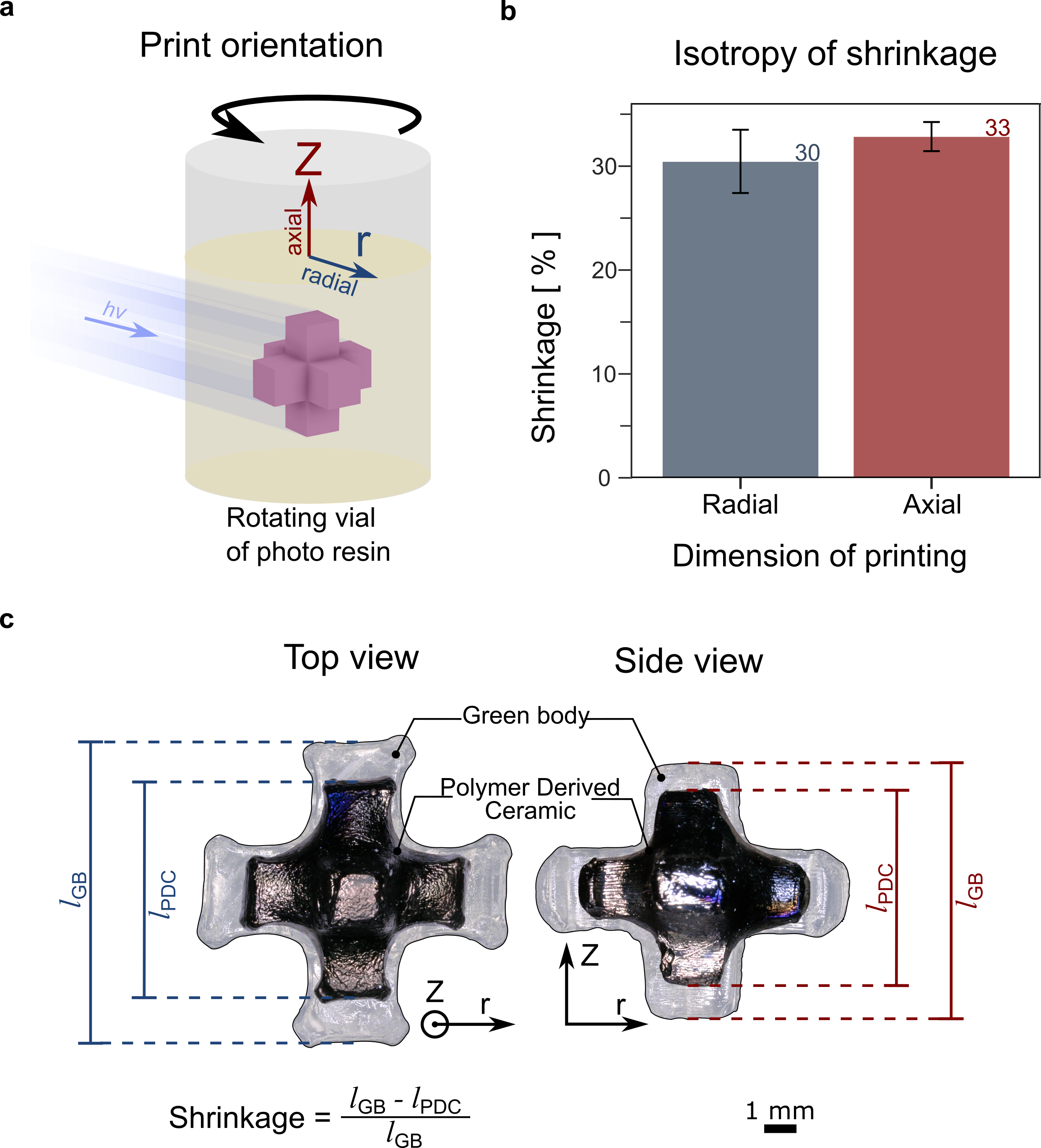}
    \caption{\textbf{Isotropy of shrinkage. a.} In tomographic volumetric additive manufacturing, the object is printed upon the simultaneous polymerization of the resin in the rotating vial. Unlike SLA or DLP, the part exhibits isotropic polymerization along cylindrical coordinates. \textbf{b.} Shrinkage along the axial and radial dimensions of prints. An unpaired t-Test shows that there is no significant difference between the shrinkage along the radial and axial dimensions. Error bars indicate a standard deviation. \textbf{c.} Overlay of the green body and polymer derived ceramic of a 3D cross.}
    \label{fig_shrinkage}
\end{figure}

    \paragraph{}Pyrolysis leads to the decomposition of organic units with an escape of volatile gases. Polymer to ceramic conversion usually happens within a temperature window of 400-800 $^{\circ}$C. Ceramic conversion is almost complete above 600 $^{\circ}$C and later there will be rearrangement of bonds to form Si-C rich and Si-O rich regimes. In our case, pyrolysis plays a key role in the stability of the final structure as both the siloxane and acrylate units have different pyrolysis profile. The acrylate starts to decompose around 375 $^{\circ}$C and care must be given at this temperature to avoid formation of bubbles and cracking of structures. Heating should be as slow as possible to allow a smooth escape of gases resulting from the decomposition of acrylate units. The pyrolysis profile is designed to allow a smooth release of volatile units. The preceramic green bodies are slowly heated to the specified temperature and a holding step of 1 hour is applied at 375 $^{\circ}$C. This is followed by a temperature ramping up to 1000 $^{\circ}$C and kept at maximum for one hour to complete the ceramic formation. This causes a large mass loss and leads to shrinkage between the green body and the polymer derived ceramic. More details are provided in the supplementary material \ref{pyrolysis} and \ref{TGA analysis}.
\paragraph{}Shrinkage poses a difficulty to fabricate functional pieces from preceramic polymers \cite{hundley2017geometricCharacOfPDCs}. Recent works measured the resulting shrinkage after pyrolysis and applied corrections to the 3D model to obtain accurate parts \cite{roopavath2019optimizationOfShrinkage}. Such corrections are more straightforward if the shrinkage is isotropic. Previous works on volumetric additive manufacturing have shown that tomographic back-projection results in isotropic, smooth polymerization, contrary to extrusion-based printing and DLP \cite{Bernal2019}. Since the green bodies are formed volumetrically, without a preferential direction, it is expected that the shrinkage is isotropic. Indeed, the pyrolyzed parts did not show significant differences in shrinkage along any direction ($p  = 6.3 \times 10^{-6}$). This allows the PDCs to keep their shape along the axial and radial dimensions of printing, as shown in Figure \ref{fig_shrinkage}. Additionally we  report a shrinkage of $31.0 \pm 1.7 \% $ and a mass loss of $54.0 \pm 0.2 \%$ from printing to pyrolysis. These results are in line with those of previous works  \cite{zanchetta2016SiOC}, \cite{hundley2017geometricCharacOfPDCs}.


\subsection{Ceramization of the polysiloxane substituted precursor}

\paragraph{} The green body is converted to the polymer derived ceramic through the pyrolytic transformation. The SPR-684 is a commercially available polysiloxane which converts to ceramic for pyrolysis temperatures tested already in the range of 1000-1400 $^{\circ}$C \cite{kaspar2012carbon,MARTINEZCRESPIERA2011913,KONSTANTINOU2020101343,MAKOWSKA2020134}. Here, both green and ceramic parts were examined by X-ray photoelectron spectroscopy (XPS). 

\begin{figure}[b!]
    \centering
    \includegraphics[width=1\textwidth]{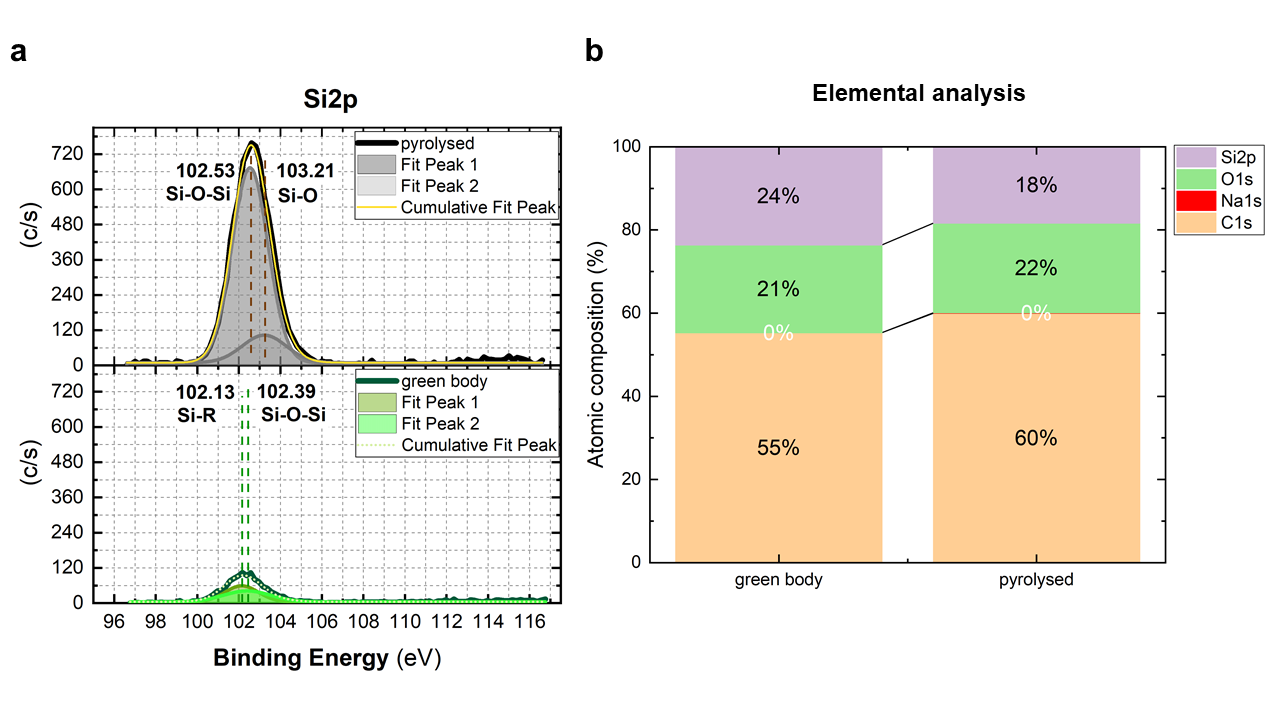}
    \caption{\textbf{a.} XPS spectra in the Si2p region of the green body and the polymer derived ceramic, \textbf{b.} Atomic compositions  \cite{chastain1992handbook}
    \label{fig3}}
\end{figure}

In Figure \ref{fig3}, the XPS (Si2p) spectra of preceramic and pyrolysed parts are presented along with respective elemental composition from the spectra. In Figure \ref{fig3}a, the two fitted peaks in the green state correspond to Si-R and Si-O-Si (from the siloxane backbone unit) \cite{chastain1992handbook,XPSdatabase} from the cross-linked preceramic gels. In the XPS spectra of pyrolysed state, a major peak at 103.21 eV appears representing Si-O bonds from amorphous SiO2 units resulting from the polymer to ceramic conversion \cite{kaspar2012carbon,MARTINEZCRESPIERA2011913,chastain1992handbook,KONSTANTINOU2020101343}. In Figure \ref{fig3}b, the elemental composition of the green and the ceramic state is presented. The amount of Na1s found is attributed to the contamination from the crucible used during pyrolysis, since in the green body no Sodium was found. The percentages of carbon and silicon are influenced by atmospheric contamination.


\subsection{Resistance of 3D printed ceramic parts}

\begin{figure}[htbp]
    \centering
    \includegraphics[width=1\textwidth]{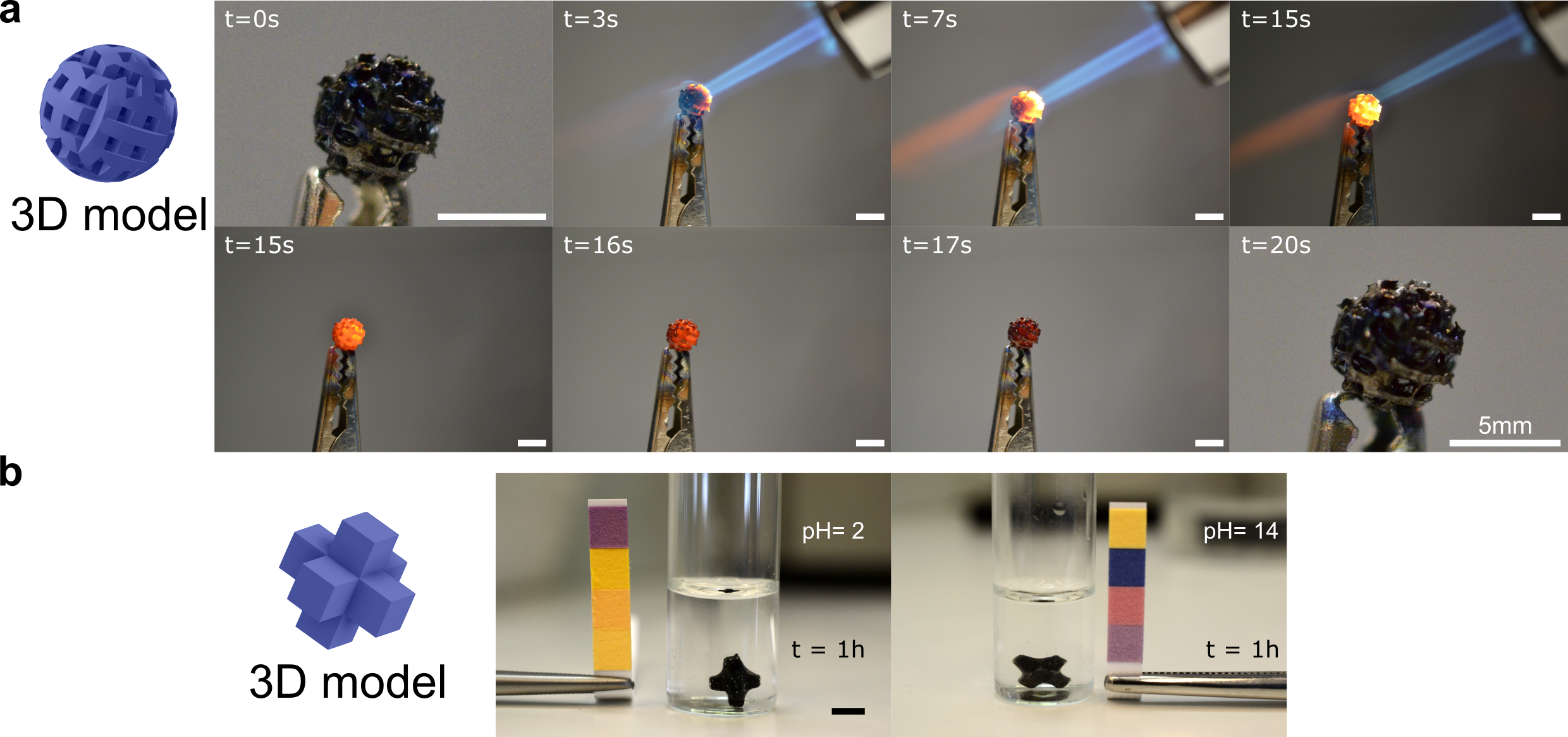}
    \caption{\textbf{Resistance of 3D printed ceramic parts. a.} Timelapse of a ceramic part being heated to incandescence with a buthane torch (T ~ 1400 $^{\circ}$C) and then let cool down. The last frame shows the part after five cycles of thermal stress. Scalebars 5 mm. \textbf{b.} Parts after being immersed for one hour in a strong acid (pH = 2) or a strong base (pH = 14) for 1 hour. Scalebars  2 mm.}
    \label{fig4}
\end{figure}

\paragraph{} 
We tested the physical and chemical properties of the fabricated PDCs. To test their thermal resistance, we exposed the parts to rapid thermal shock cycles of 15 seconds heating up under the flame of a butane torch and 10 seconds of cooling down. The temperature of the flame ($\textit{T} \approx $1400 $^{\circ}$C) is higher than the pyrolysis temperature. Figure \ref{fig4}a shows a time-lapse sequence of a spherical woodpile under its fifth thermal stress cycle. The first and last frames of the time-lapse show that the part retained its shape and did not crack, even withstanding the stress induced by the holding clamp.
To assess the chemical inertness of the parts, we submerged them for 1 hour in aqueous corrosive baths. Figure \ref{fig4}b shows a 3D cross PDC sitting in a HCl solution of pH = 2 on the left and a 3D cross PDC sitting in a KOH solution of pH = 14 on the right (see Figure \ref{pH} in appendix). Both parts retained their mass (within 0.1 mg on a precision scale). This demonstrates that they are very resistant to high temperatures, rapid heating and cooling for several cycles, and to corrosion. 

\subsection{Examples of 3D volumetric printed ceramic parts}
\begin{figure}[t!]
    \centering
    \includegraphics[width=150mm]{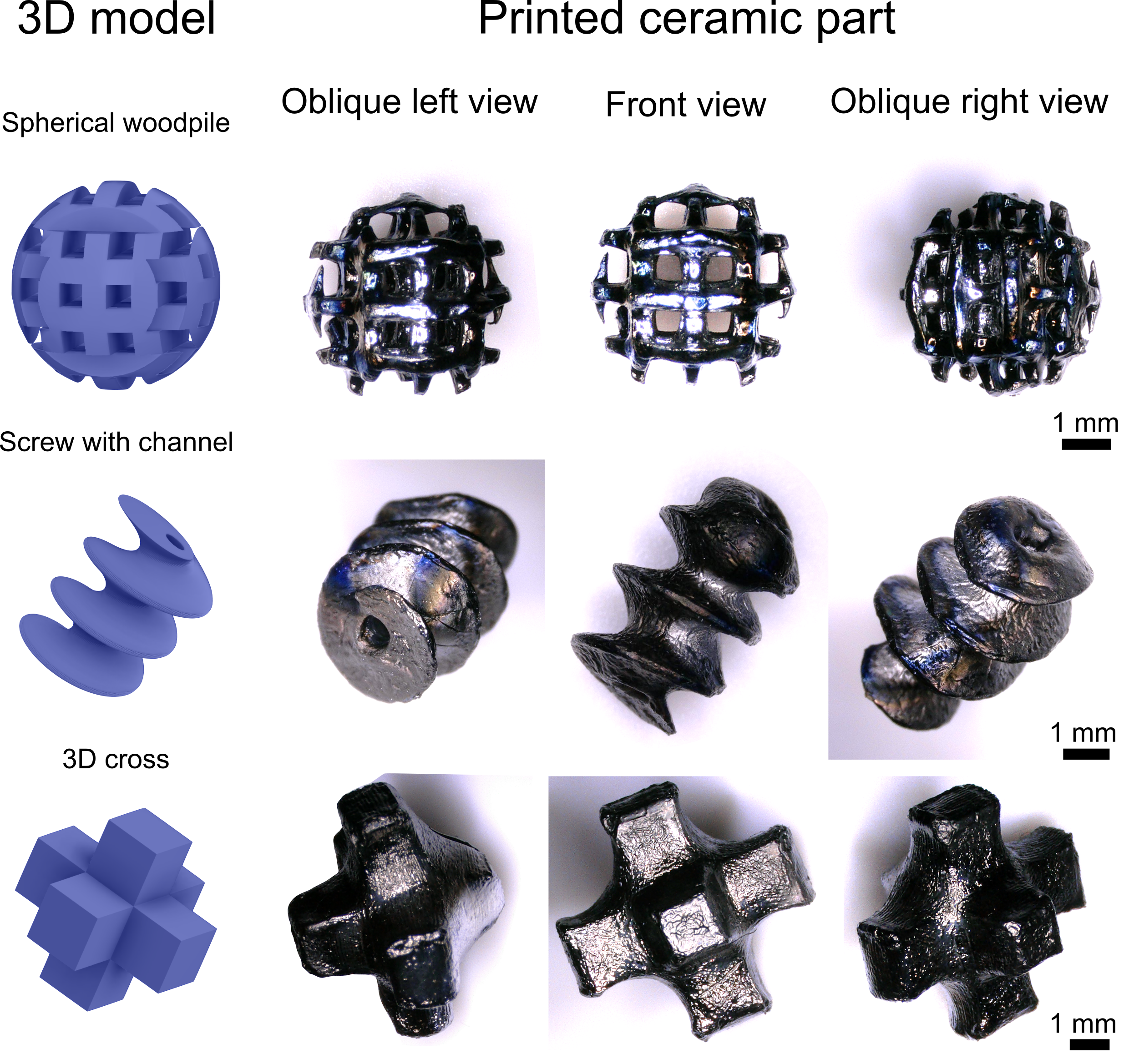}
    \caption{\textbf{Examples of 3D volumetric printed ceramic parts.} Top one is a spherical woodpile. Middle one is a screw with a channel. Bottom one is a 3D cross. Scalebar  1 mm.}
    \label{fig5}
\end{figure}

Volumetric additive manufacturing enables the fabrication of microcomponents with unique geometries which are very difficult to obtain with other AM technologies. This is illustrated in Figure \ref{fig5}. The top row depicts a 10-level spherical woodpile structure with rods of 1mm in width and height spaced by 1mm, cropped in the shape of a ball. After pyrolysis, the 31$\%$ shrinkage brings the size of the ball down to $\approx$ 6 mm diameter with features of 700 µm. The shape was successfully pyrolyzed without bubble formation, deformation or cracks. This shows that relatively large objects can be produced, as long as escape routes are added for the gases in the form of empty channels.
The second row presents a screw with 3 full revolutions, 7.2 mm in length, an outer diameter of 6 mm, and  with a channel of diameter 1 mm going down its center (Figure \ref{uCT}). After pyrolysis, the channel diameter was reduced to 700 µm and length of the screw reduced to around 5.75 mm. Some air bubbles were trapped within the PDCs during pyrolysis, thus slightly deforming the part. This shape shows that it is possible to have features on the outside of the part while having a void channel in the center. 
The third shape is a 3D cross with arms of dimensions 3 mm $\times$ 3 mm $\times$ 3 mm. After pyrolysis, those dimensions got reduced to 2mm $\times$ 2mm $\times$ 2mm. Again this shape deformed because of a bubble in its center. The bubble is slightly off center, which gives it this asymmetrical deformation, with one side unaffected and the other with a swollen arm.

\section{Discussion and conclusion}


\justifying

\paragraph{}In this work, we demonstrated the fabrication of low-shrinkage, isotropic, fully dense and crack-free PDCs from volumetric additive manufacturing. We formulated a transparent and highly viscous preceramic resin which is added in a rotating, transparent cylindrical vial photocured with a laser source at 405 nm and using tomographic back-projection. Parts with complex shapes were successfully fabricated and the green bodies were converted to ceramic parts through a pyrolysis step. We reported qualitatively on the density and the smoothness of the parts at the microscopic scale by micro Computed Tomography and Scanning Electron Microscopy. Based on these experimental measurements, we estimated the isotropy of shrinkage and the mass loss. Lastly, we validated the polymer to ceramic conversion by X-ray photoelectron spectroscopy and confirmed the high-resistance to heat and acidic or basic environments.
Ceramics are popular candidates for the fabrication of prototypes with exceptional properties, but traditional AM techniques impose long building time because of their inherent layer-by-layer process. Consequently, both steps of the 3D printing and the pyrolysis are time consuming, which stands in the way of rapidly optimizing a ceramic prototype until specific requirements are met (accuracy, precision and tolerance). Volumetric printing is an emerging 3D printing technology that drastically accelerates the 3D printing step, leads to isotropic shrinkage properties, and opens up a different range of materials and geometries for use in prototyping ceramics. Future work on the accuracy assessment step could provide even faster cycles to optimize prototypes.

\section{Experimental Section}

\justifying

\subsection*{Preceramic resin}
The preceramic resin was prepared by combining a polysiloxane substituted precursor (SPR 684, Starfire Systems, USA) with 1,4-butanediol diacrylate (BDDA) as a crosslinker (1070-70-8, Sigma Aldrich, USA), and Diphenyl (2,4,6-trimethylbenzoyl) phosphine oxide (TPO) as a photoinitiator (75980-60-8, Sigma Aldrich). The resin preparation consisted of 85 wt\% polysiloxane, 15 wt\% BDDA and 2mM TPO (0.063 wt\%).

To produce the resin, a solution of diluted TPO in BDDA is prepared to a concentration of 30 mg/mL which is then vortexed. Following this, the polysiloxane precursor is combined with the TPO in the BBDA solution. The components are then simultaneously mixed and degassed  using a planetary mixer (Mazerustar KK-250SE, Kurabo, Japan). Finally, the resin is poured into glass vials of 16.5 mm, which are used for printing.

\subsection*{Viscosity}
The viscosity of the resin was measured using a rheometer (MCR 102, Anton-Paar, Austria) with a (25 mm) parallel plate and a gap of (350 µm) at a shear rate of 0.108 Pa

\subsection*{Tomographic volumetric additive manufacturing}
\paragraph{3D printer}
A custom volumetric 3D printer\cite{Loterie2020} is used for this work. In it, the light from 4 laser diodes ($\lambda$ = 405 nm) is coupled into a multimode fiber. Then, the beam is expanded to fill a Digital Micromirror Device, DMD, (VIS-7000, Vialux, Germany). The patterns for the tomographic reconstruction of the part are displayed on the DMD and projected by a pair of achromatic lenses with  focal lengths of f1 = 100 mm (AC254-100-A-ML, Thorlabs, USA) and f2 =  250 mm (ACT508-250-A-ML, Thorlabs) onto the rotating cylindrical glass vial containing the photopolymer. The resin vial is dipped in a refractive index matching bath with square footprint to remove the lensing effects of the round vial. A camera records the progress of the print in the vial by imaging the sample with an orthogonal 671 nm expanded laser beam.

\paragraph{Tomographic backprojection} 
The patterns for the tomographic backprojection are computed following the algorithm described in \cite{Loterie2020}.  To do so, the STL file of the 3D part is voxelized using a python script. Then the Radon transform of this digital object is calculated over \(\pi\) rad around one of its axes. The Fourier transform of these patterns is taken and a ramp filter is applied  to compensate for the over sampling of the low spatial frequencies. Finally, the inverse Fourier transform of these patterns is calculated. Additional corrections for attenuation from the photoinitiator were also performed following \cite{madrid2021scattering} .

\subsection*{Postprocessing of prints}
After printing, the parts are recovered from the glass vials and dipped into a toluene bath, which is manually stirred for 5 minutes until all uncured resin dissolves. The parts are then placed in a bath of isopropyl alcohol (IPA) to dilute the toluene and stop the solving action on the parts. The bath is manually stirred again for about 1 minute. The parts are then left to dry in air at room temperature until all IPA fully evaporated, leaving them free of any unpolymerised resin. 

The parts are then postcured in a UV curing station (FormCure, Formlabs, USA) for 1 hour at room temperature. After this, all remaining photoinitiator has been consumed, but the surface of the parts is still sticky. To remove the stickiness, the green bodies are submerged in a concentrated solution of TPO in IPA (10 mg/mL) and left for 1 hour to allow TPO to diffuse inside. The bath with the parts is then placed for 15 minutes in the UV curing station. Then, the parts are removed from the bath and postcured one last time for 45 minutes in the curing station. After that, the green bodies are placed in an oven for 24 hours at 80 $^{\circ}$C to remove most of the solvent soaked into the part. 

\subsection*{Pyrolysis}
The rinsed, postcured, and aged green bodies are then pyrolysed in an alumina tube furnace (STF 15/450, Carbolite Gero, Germany) under argon atmosphere following the temperature profile described and explained in the supplementary materials (\ref{pyrolysis}). The pyrolysis peak temperature was set to T = 1000 $^{\circ}$C, for a dwell time of 1 hour, and a total cycle duration of 48 hours (The actual pyrolysis temperature profile is shown in the appendix: Figure \ref{pyrolysis}.

\subsection*{Imaging}
\paragraph{µCT Imaging}
Printed objects were imaged with voxel sizes of 10 µm $\times$ 10 µm $\times$ 10 µm  under a 160 kV X-ray transmission tomograph (Hamamatsu, Japan). 3D visualizations of the pieces were obtained using Fiji-ImageJ \cite{schindelin2012fiji}.

\paragraph{Photographic Imaging}
Green bodies and pyrolyzed parts were imaged with a DSLR camera (D3100, Nikon, Japan) with a \textit{f}=2.8 macro lens (AF-S Micro Nikkor 40 mm, Nikon) , and a digital microscope (VHX-5000. Keyence, USA).

\paragraph{Gold deposition and SEM}

The sputtering machine Alliance-Concept DP650 was used for the deposition of a thin gold layer (20nm) on the samples. Following that, the samples were transferred to the Scanning Electron Microscope (SEM LEO 1550) and were inserted into the chamber vacuum for imaging the surface from low to high magnification.

\subsection*{Thermal stress resistance}
To show resistance of the ceramic parts to high temperature, a butane torch was heating the ceramic parts  (T $\approx$ 1400 $^{\circ}$C) for some seconds until they became incandescent and then let cool down. A typical thermal stress cycle was 20 seconds. The spherical woodpile shown in Figure \ref{fig4}  was subjected to 5 thermal stress cycles.

\subsection*{Chemical resistance}
Parts were dipped into vials containing aqueous solutions of HCl and KOH solutions (pH = 2, and pH = 14 respectively) for one hour and photographed at the beginning and the end of the experiment.

\subsection*{X-ray photoelectron Spectroscopy}
The polymer to ceramic conversion is validated by X-Ray Photoelectron Spectroscopy (XPS) measurements. The bonds of Silicon that occur demonstrate that SiOC ceramic is formed.

\subsection*{Shrinkage}
To compare the differences between shrinkage along the axial and radial dimensions of the prints, a set of lengths were measured on green bodies and their corresponding PDCs ($n_{parts} = 7$, $n_{measurements} = 14$), most of them woodpiles. Flat geometries were chosen because they reduced ambiguity in measuring lengths, as shown in Figure \ref{shrinkage}. 

\subsection*{Statistical analysis}
Statistical analysis of the isotropy of shrinkage was conducted by running a two-tailed t-Test assuming unequal variances (Welch test, $\alpha = 0.05$) on Microsoft Excel.

\newpage

\medskip
\textbf{Supporting Information} \par 
Supporting Information is available from the Wiley Online Library or from the author.

\medskip
\textbf{Acknowledgements} \par 
\justifying

MK., GK, and JMW contributed equally to this work. This work was supported by Strategic Focus Area (SFA)- Advanced Manufacturing for the Ceramic X.0 –High-precision micro-manufacturing of ceramics.
The authors thank Lukas Riemer (Group for Ferroelectrics and Functional Oxides - EPFL) and Lorenz Hagelucken and Juergen Brugger (Microsystems Laboratory 1 - EPFL) for their help with the pyrolisis and useful discussions. The authors also acknowledge Gary Perrenoud and the PIXE Facility at EPFL for the $\mu$CT scans of the parts. The authors acknowldge Dr. Mounir Driss Mensi  (X-Ray Diffraction and Surface Analytics Platform - EPFL),  Dr. Pierre Mettraux (Molecular and Hybrid Materials Characterization Center - EPFL; \textit{in memoriam}) and  Dr. Nikolaos Nianias for their help with  X-ray Photoelectron Spectroscopy. 
We thank the large number of open-source and related projects that facilitated this work, including FreeCAD, ImageJ, Python, Anaconda, Binvox, and Inkscape.

\medskip
\printbibliography

@article{Loterie2020,
  doi = {10.1038/s41467-020-14630-4},
  url = {https://doi.org/10.1038/s41467-020-14630-4},
  year = {2020},
  month = feb,
  publisher = {Springer Science and Business Media {LLC}},
  volume = {11},
  number = {1},
  author = {Damien Loterie and Paul Delrot and Christophe Moser},
  title = {High-resolution tomographic volumetric additive manufacturing},
  journal = {Nature Communications}
}

@article{Kelly2019,
  doi = {10.1126/science.aau7114},
  url = {https://doi.org/10.1126/science.aau7114},
  year = {2019},
  month = jan,
  publisher = {American Association for the Advancement of Science ({AAAS})},
  volume = {363},
  number = {6431},
  pages = {1075--1079},
  author = {Brett E. Kelly and Indrasen Bhattacharya and Hossein Heidari and Maxim Shusteff and Christopher M. Spadaccini and Hayden K. Taylor},
  title = {Volumetric additive manufacturing via tomographic reconstruction},
  journal = {Science}
}

@article{madrid2021scattering,
  title={Light-based Volumetric Additive Manufacturing in Scattering Resins},
  author={Madrid-Wolff, Jorge and Boniface, Antoine and Loterie, Damien and Delrot, Paul and Moser, Christophe},
  journal={arXiv preprint arXiv:2105.14952},
  year={2021}
}

@article{regehly2020xolography,
  title={Xolography for linear volumetric 3D printing},
  author={Regehly, Martin and Garmshausen, Yves and Reuter, Marcus and K{\"o}nig, Niklas F and Israel, Eric and Kelly, Damien P and Chou, Chun-Yu and Koch, Klaas and Asfari, Baraa and Hecht, Stefan},
  journal={Nature},
  volume={588},
  number={7839},
  pages={620--624},
  year={2020},
  publisher={Nature Publishing Group}
}

@article{cook2020ThiolEne,
  title={Highly Tunable Thiol-Ene Photoresins for Volumetric Additive Manufacturing},
  author={Cook, Caitlyn C and Fong, Erika J and Schwartz, Johanna J and Porcincula, Dominique H and Kaczmarek, Allison C and Oakdale, James S and Moran, Bryan D and Champley, Kyle M and Rackson, Charles M and Muralidharan, Archish and others},
  journal={Advanced Materials},
  volume={32},
  number={47},
  pages={2003376},
  year={2020},
  publisher={Wiley Online Library}
}

@misc{rackson2021StriationFreeVAM,
  author       = {Rackson, Charles and McLeod, Robert},
  title        = {Striation-Free Volumetric Additive Manufacturing},
  howpublished = {Volumetric Additive Manufacturing Workshop},
  month        = {August},
  year         = {2021},
}

@article{shoji1999optically,
  title={Optically-induced growth of fiber patterns into a photopolymerizable resin},
  author={Shoji, Satoru and Kawata, Satoshi},
  journal={Applied physics letters},
  volume={75},
  number={5},
  pages={737--739},
  year={1999},
  publisher={American Institute of Physics}
}

@article{ligon2014OxygenInhibition,
  title={Strategies to reduce oxygen inhibition in photoinduced polymerization},
  author={Ligon, Samuel Clark and Husar, Branislav and Wutzel, Harald and Holman, Richard and Liska, Robert},
  journal={Chemical reviews},
  volume={114},
  number={1},
  pages={557--589},
  year={2014},
  publisher={ACS Publications}
}

@article{flory1941molecular,
  title={Molecular size distribution in three dimensional polymers. I. Gelation1},
  author={Flory, Paul J},
  journal={Journal of the American Chemical Society},
  volume={63},
  number={11},
  pages={3083--3090},
  year={1941},
  publisher={ACS Publications}
}

@article{zocca2015additive,
  title={Additive manufacturing of ceramics: issues, potentialities, and opportunities},
  author={Zocca, Andrea and Colombo, Paolo and Gomes, Cynthia M and G{\"u}nster, Jens},
  journal={Journal of the American Ceramic Society},
  volume={98},
  number={7},
  pages={1983--2001},
  year={2015},
  publisher={Wiley Online Library}
}

@article{chen20193d,
  title={3D printing of ceramics: A review},
  author={Chen, Zhangwei and Li, Ziyong and Li, Junjie and Liu, Chengbo and Lao, Changshi and Fu, Yuelong and Liu, Changyong and Li, Yang and Wang, Pei and He, Yi},
  journal={Journal of the European Ceramic Society},
  volume={39},
  number={4},
  pages={661--687},
  year={2019},
  publisher={Elsevier}
}

@book{chartier2013handbook,
  title={Handbook of Advanced Ceramics: Chapter 6.5. Rapid Prototyping of Ceramics},
  author={Chartier, T and Badev, A},
  year={2013},
  publisher={Elsevier Inc. Chapters}
}

@book{somiya2013handbook,
  title={Handbook of advanced ceramics: materials, applications, processing, and properties},
  author={Somiya, Shigeyuki},
  year={2013},
  publisher={Academic press}
}

@article{chen2021dense,
  title={Dense ceramics with complex shape fabricated by 3D printing: A review},
  author={Chen, Zhe and Sun, Xiaohong and Shang, Yunpeng and Xiong, Kunzhou and Xu, Zhongkai and Guo, Ruisong and Cai, Shu and Zheng, Chunming},
  journal={Journal of Advanced Ceramics},
  pages={1--24},
  year={2021},
  publisher={Springer}
}

@article{pelza2021additive,
  title={Additive Manufacturing of Structural Ceramics: A Historical Perspective},
  author={Pelza, Joshua S and Ku, Nicholas and Meyers, Marc A and Vargas-Gonzalez, Lionel R},
  journal={Journal of Materials Research and Technology},
  year={2021},
  publisher={Elsevier}
}

@article{he2021progress,
  title={Progress and challenges towards additive manufacturing of SiC ceramic},
  author={He, Rujie and Zhou, Niping and Zhang, Keqiang and Zhang, Xueqin and Zhang, Lu and Wang, Wenqing and Fang, Daining},
  journal={Journal of Advanced Ceramics},
  volume={10},
  number={4},
  pages={637--674},
  year={2021},
  publisher={Springer}
}

@article{greil2002advanced,
  title={Advanced engineering ceramics},
  author={Greil, Peter},
  journal={Advanced Engineering Materials},
  volume={4},
  number={5},
  pages={247--254},
  year={2002},
  publisher={Wiley Online Library}
}

@article{liu20203d,
  title={3D printing of ceramic cellular structures for potential nuclear fusion application},
  author={Liu, Yu and Chen, Zhangwei and Li, Junjie and Gong, Baoping and Wang, Long and Lao, Changshi and Wang, Pei and Liu, Changyong and Feng, Yongjin and Wang, Xiaoyu},
  journal={Additive Manufacturing},
  volume={35},
  pages={101348},
  year={2020},
  publisher={Elsevier}
}

@article{hundley2017geometricCharacOfPDCs,
  title={Geometric characterization of additively manufactured polymer derived ceramics},
  author={Hundley, Jacob M and Eckel, Zak C and Schueller, Emily and Cante, Kenneth and Biesboer, Scott M and Yahata, Brennan D and Schaedler, Tobias A},
  journal={Additive Manufacturing},
  volume={18},
  pages={95--102},
  year={2017},
  publisher={Elsevier}
}

@article{park2009fabrication,
  title={Fabrication of three-dimensional SiC ceramic microstructures with near-zero shrinkage via dual crosslinking induced stereolithography},
  author={Park, Sungjune and Lee, Dong-Hoon and Ryoo, Hyang-Im and Lim, Tae-Woo and Yang, Dong-Yol and Kim, Dong-Pyo},
  journal={Chemical communications},
  number={32},
  pages={4880--4882},
  year={2009},
  publisher={Royal Society of Chemistry}
}

@article{eckel2016additive,
  title={Additive manufacturing of polymer-derived ceramics},
  author={Eckel, Zak C and Zhou, Chaoyin and Martin, John H and Jacobsen, Alan J and Carter, William B and Schaedler, Tobias A},
  journal={Science},
  volume={351},
  number={6268},
  pages={58--62},
  year={2016},
  publisher={American Association for the Advancement of Science}
}

@article{schmidt2018digital,
  title={Digital light processing of ceramic components from polysiloxanes},
  author={Schmidt, Johanna and Colombo, Paolo},
  journal={Journal of the European Ceramic Society},
  volume={38},
  number={1},
  pages={57--66},
  year={2018},
  publisher={Elsevier}
}

@article{schmidt2019multiscale,
  title={Multiscale ceramic components from preceramic polymers by hybridization of vat polymerization-based technologies},
  author={Schmidt, Johanna and Brigo, Laura and Gandin, Alessandro and Schwentenwein, Martin and Colombo, Paolo and Brusatin, Giovanna},
  journal={Additive Manufacturing},
  volume={30},
  pages={100913},
  year={2019},
  publisher={Elsevier}
}

@article{zanchetta2016SiOC,
author = {Zanchetta, Erika and Cattaldo, Marco and Franchin, Giorgia and Schwentenwein, Martin and Homa, Johannes and Brusatin, Giovanna and Colombo, Paolo},
title = {Stereolithography of SiOC Ceramic Microcomponents},
journal = {Advanced Materials},
volume = {28},
number = {2},
pages = {370-376},
doi = {https://doi.org/10.1002/adma.201503470},
url = {https://onlinelibrary.wiley.com/doi/abs/10.1002/adma.201503470},
eprint = {https://onlinelibrary.wiley.com/doi/pdf/10.1002/adma.201503470},
year = {2016}
}

@inproceedings{schwentenwein2014lithography,
  title={Lithography-based ceramic manufacturing: A novel technique for additive manufacturing of high-performance ceramics},
  author={Schwentenwein, Martin and Schneider, Peter and Homa, Johannes},
  booktitle={Advances in Science and Technology},
  volume={88},
  pages={60--64},
  year={2014},
  organization={Trans Tech Publ}
}

@article{roopavath2019optimizationOfShrinkage,
  title={Optimization of extrusion based ceramic 3D printing process for complex bony designs},
  author={Roopavath, Uday Kiran and Malferrari, Sara and Van Haver, Annemieke and Verstreken, Frederik and Rath, Subha Narayan and Kalaskar, Deepak M},
  journal={Materials \& Design},
  volume={162},
  pages={263--270},
  year={2019},
  publisher={Elsevier}
}

@article{KONSTANTINOU2020101343,
title = {Additive micro-manufacturing of crack-free PDCs by two-photon polymerization of a single, low-shrinkage preceramic resin},
journal = {Additive Manufacturing},
volume = {35},
pages = {101343},
year = {2020},
issn = {2214-8604},
doi = {https://doi.org/10.1016/j.addma.2020.101343},
url = {https://www.sciencedirect.com/science/article/pii/S2214860420307156},
author = {Georgia Konstantinou and Eirini Kakkava and Lorenz Hagelüken and Pradeep Vallachira {Warriam Sasikumar} and Jieping Wang and Malgorzata Grazyna Makowska and Gurdial Blugan and Nikolaos Nianias and Federica Marone and Helena {Van Swygenhoven} and Jürgen Brugger and Demetri Psaltis and Christophe Moser}
}

@article{brigo20183d,
  title={3D nanofabrication of SiOC ceramic structures},
  author={Brigo, Laura and Schmidt, Johanna Eva Maria and Gandin, Alessandro and Michieli, Niccol{\`o} and Colombo, Paolo and Brusatin, Giovanna},
  journal={Advanced Science},
  volume={5},
  number={12},
  pages={1800937},
  year={2018},
  publisher={Wiley Online Library}
}

@article{MAKOWSKA2020134,
title = {Cracks, porosity and microstructure of Ti modified polymer-derived SiOC revealed by absorption-, XRD- and XRF-contrast 2D and 3D imaging},
journal = {Acta Materialia},
volume = {198},
pages = {134-144},
year = {2020},
issn = {1359-6454},
doi = {https://doi.org/10.1016/j.actamat.2020.07.067},
url = {https://www.sciencedirect.com/science/article/pii/S135964542030584X},
author = {Małgorzata Makowska and Pradeep Vallachira Warriam Sasikumar and Lorenz Hagelüken and Dario F. Sanchez and Nicola Casati and Federica Marone and Gurdial Blugan and Jürgen Brugger and Helena {Van Swygenhoven}}
}

@article{deFaoite2011,
  doi = {10.1007/s10853-011-6140-1},
  url = {https://doi.org/10.1007/s10853-011-6140-1},
  year = {2011},
  month = dec,
  publisher = {Springer Science and Business Media {LLC}},
  volume = {47},
  number = {10},
  pages = {4211--4235},
  author = {Daith{\'{\i}} de Faoite and David J. Browne and Franklin R. Chang-D{\'{\i}}az and Kenneth T. Stanton},
  title = {A review of the processing,  composition,  and temperature-dependent mechanical and thermal properties of dielectric technical ceramics},
  journal = {Journal of Materials Science}
}

@article{Fu2019,
  doi = {10.1007/s40145-019-0335-3},
  url = {https://doi.org/10.1007/s40145-019-0335-3},
  year = {2019},
  month = dec,
  publisher = {Springer Science and Business Media {LLC}},
  volume = {8},
  number = {4},
  pages = {457--478},
  author = {Shengyang Fu and Min Zhu and Yufang Zhu},
  title = {Organosilicon polymer-derived ceramics: An overview},
  journal = {Journal of Advanced Ceramics}
}

@article{He2018,
  doi = {10.1016/j.ceramint.2017.11.135},
  url = {https://doi.org/10.1016/j.ceramint.2017.11.135},
  year = {2018},
  month = feb,
  publisher = {Elsevier {BV}},
  volume = {44},
  number = {3},
  pages = {3412--3416},
  author = {Rongxuan He and Wei Liu and Ziwei Wu and Di An and Meipeng Huang and Haidong Wu and Qiangguo Jiang and Xuanrong Ji and Shanghua Wu and Zhipeng Xie},
  title = {Fabrication of complex-shaped zirconia ceramic parts via a {DLP}- stereolithography-based 3D printing method},
  journal = {Ceramics International}
}

@article{JANSSEN20081369,
title = {Tailor-made ceramic-based components—Advantages by reactive processing and advanced shaping techniques},
journal = {Journal of the European Ceramic Society},
volume = {28},
number = {7},
pages = {1369-1379},
year = {2008},
note = {Developments in Ceramic Science and Engineering: the last 50 years. A meeting in celebration of Professor Sir Richard Brook's 70th Birthday},
issn = {0955-2219},
doi = {https://doi.org/10.1016/j.jeurceramsoc.2007.12.022},
url = {https://www.sciencedirect.com/science/article/pii/S0955221907006115},
author = {Rolf Janssen and Sven Scheppokat and Nils Claussen}
}

@article{MARTINEZCRESPIERA2011913,
title = {Pressureless synthesis of fully dense and crack-free SiOC bulk ceramics via photo-crosslinking and pyrolysis of a polysiloxane},
journal = {Journal of the European Ceramic Society},
volume = {31},
number = {5},
pages = {913-919},
year = {2011},
issn = {0955-2219},
doi = {https://doi.org/10.1016/j.jeurceramsoc.2010.11.019},
url = {https://www.sciencedirect.com/science/article/pii/S0955221910005388},
author = {Sandra Martínez-Crespiera and Emanuel Ionescu and Hans-Joachim Kleebe and Ralf Riedel}
}

@book{lalevee2015dyes,
  title={Dyes and chromophores in polymer science},
  author={Lalev{\'e}e, Jacques and Fouassier, Jean-Pierre},
  year={2015},
  publisher={John Wiley \& Sons}
}

@article{eibel2018choosing,
  title={Choosing the ideal photoinitiator for free radical photopolymerizations: Predictions based on simulations using established data},
  author={Eibel, Anna and Fast, David E and Gescheidt, Georg},
  journal={Polymer Chemistry},
  volume={9},
  number={41},
  pages={5107--5115},
  year={2018},
  publisher={Royal Society of Chemistry}
}

@article{slavin2012cobalt,
  title={Cobalt-Catalyzed Chain Transfer Polymerization: A Review},
  author={Slavin, S and McEwan, K and Haddleton, DM},
  journal={Polymer Science: A Comprehensive Reference, 10 Volume Set},
  pages={249--275},
  year={2012},
  publisher={Elsevier}
}

@article{LU2016411,
title = {Fundamental understanding of water vapor effect on SiOC evolution during pyrolysis},
journal = {Journal of the European Ceramic Society},
volume = {36},
number = {3},
pages = {411-422},
year = {2016},
issn = {0955-2219},
doi = {https://doi.org/10.1016/j.jeurceramsoc.2015.11.003},
url = {https://www.sciencedirect.com/science/article/pii/S0955221915302168},
author = {Kathy Lu and Jiake Li}
}

@article{SASIKUMAR201820961,
title = {Polymer derived silicon oxycarbide ceramic monoliths: Microstructure development and associated materials properties},
journal = {Ceramics International},
volume = {44},
number = {17},
pages = {20961-20967},
year = {2018},
issn = {0272-8842},
doi = {https://doi.org/10.1016/j.ceramint.2018.08.102},
url = {https://www.sciencedirect.com/science/article/pii/S0272884218321539},
author = {Pradeep Vallachira Warriam Sasikumar and Gurdial Blugan and Nicola Casati and Eirini Kakkava and Giulia Panusa and Demetri Psaltis and Jakob Kuebler}
}

@article{Xu2019,
  doi = {10.1002/slct.201900993},
  url = {https://doi.org/10.1002/slct.201900993},
  year = {2019},
  month = jun,
  publisher = {Wiley},
  volume = {4},
  number = {23},
  pages = {6862--6869},
  author = {Xiaobo Xu and Peiying Li and Chunhua Ge and Weifang Han and Di Zhao and Xiangdong Zhang},
  title = {3D Printing of Complex-type {SiOC} Ceramics Derived From Liquid Photosensitive Resin},
  journal = {{ChemistrySelect}}
}

@article{he2020polymer,
  title={Polymer-derived SiOC ceramic lattice with thick struts prepared by digital light processing},
  author={He, Chong and Ma, Cong and Li, Xilu and Yan, Liwen and Hou, Feng and Liu, Jiachen and Guo, Anran},
  journal={Additive Manufacturing},
  volume={35},
  pages={101366},
  year={2020},
  publisher={Elsevier}
}

@article{GROSSENBACHER2015623,
title = {On the micrometre precise mould filling of liquid polymer derived ceramic precursor for 300-µm-thick high aspect ratio ceramic MEMS},
journal = {Ceramics International},
volume = {41},
number = {1, Part A},
pages = {623-629},
year = {2015},
issn = {0272-8842},
doi = {https://doi.org/10.1016/j.ceramint.2014.08.112},
url = {https://www.sciencedirect.com/science/article/pii/S0272884214013492},
author = {Jonas Grossenbacher and Maurizio R. Gullo and Vadym Bakumov and Gurdial Blugan and Jakob Kuebler and Juergen Brugger}
}

@article{ZHANG1995729,
title = {Injection moulding of silicon carbide using an organic vehicle based on a preceramic polymer},
journal = {Journal of the European Ceramic Society},
volume = {15},
number = {8},
pages = {729-734},
year = {1995},
issn = {0955-2219},
doi = {https://doi.org/10.1016/0955-2219(95)00049-Z},
url = {https://www.sciencedirect.com/science/article/pii/095522199500049Z},
author = {T. Zhang and J.R.G. Evans and J. Woodthorpe}
}

@article{Bernal2019,
  doi = {10.1002/adma.201904209},
  url = {https://doi.org/10.1002/adma.201904209},
  year = {2019},
  month = aug,
  publisher = {Wiley},
  volume = {31},
  number = {42},
  pages = {1904209},
  author = {Paulina Nu{\~{n}}ez Bernal and Paul Delrot and Damien Loterie and Yang Li and Jos Malda and Christophe Moser and Riccardo Levato},
  title = {Volumetric Bioprinting of Complex Living-Tissue Constructs within Seconds},
  journal = {Advanced Materials}
}

@article{Jiang2018,
  doi = {10.3390/jmmp2040064},
  url = {https://doi.org/10.3390/jmmp2040064},
  year = {2018},
  month = sep,
  publisher = {{MDPI} {AG}},
  volume = {2},
  number = {4},
  pages = {64},
  author = {Jingchao Jiang and Xun Xu and Jonathan Stringer},
  title = {Support Structures for Additive Manufacturing: A Review},
  journal = {Journal of Manufacturing and Materials Processing}
}

@article{chastain1992handbook,
  title={Handbook of X-ray photoelectron spectroscopy},
  author={Chastain, Jill and King Jr, Roger C},
  journal={Perkin-Elmer Corporation},
  volume={40},
  pages={221},
  year={1992}
}

@online{XPSdatabase,
 author = {NIST},
 title = {xps database},
 year = 2021,
 url = {https://srdata.nist.gov/xps/EngElmSrchQuery.aspx?EType=PE&CSOpt=Retri_ex_dat&Elm=Si},
 urldate = {2021-09-7}}

@article{kaspar2012carbon,
  title={Carbon-rich SiOC anodes for lithium-ion batteries: Part II. Role of thermal cross-linking},
  author={Kaspar, Jan and Graczyk-Zajac, Magdalena and Riedel, Ralf},
  journal={Solid State Ionics},
  volume={225},
  pages={527--531},
  year={2012},
  publisher={Elsevier}
}

@article{Colombo2010,
  doi = {10.1111/j.1551-2916.2010.03876.x},
  url = {https://doi.org/10.1111/j.1551-2916.2010.03876.x},
  year = {2010},
  month = jun,
  publisher = {Wiley},
  author = {Paolo Colombo and Gabriela Mera and Ralf Riedel and Gian Domenico Sorar{\`{u}}},
  title = {Polymer-Derived Ceramics: 40 Years of Research and Innovation in Advanced Ceramics},
  journal = {Journal of the American Ceramic Society}
}

@article{mera2013polymer,
  title={Polymer-derived SiCN and SiOC ceramics--structure and energetics at the nanoscale},
  author={Mera, Gabriela and Navrotsky, Alexandra and Sen, Sabyasachi and Kleebe, Hans-Joachim and Riedel, Ralf},
  journal={Journal of Materials Chemistry A},
  volume={1},
  number={12},
  pages={3826--3836},
  year={2013},
  publisher={Royal Society of Chemistry}
}

@article{schindelin2012fiji,
  title={Fiji: an open-source platform for biological-image analysis},
  author={Schindelin, Johannes and Arganda-Carreras, Ignacio and Frise, Erwin and Kaynig, Verena and Longair, Mark and Pietzsch, Tobias and Preibisch, Stephan and Rueden, Curtis and Saalfeld, Stephan and Schmid, Benjamin and others},
  journal={Nature methods},
  volume={9},
  number={7},
  pages={676--682},
  year={2012},
  publisher={Nature Publishing Group}
}

\beginsupplement
\section*{Supplementary Information}

\justifying

\section*{Optical, rheological, and mechanical properties of the preceramic resin}

\subsection*{Optical properties}

\begin{figure}[H]
    \centering
    \includegraphics[width=1\textwidth]{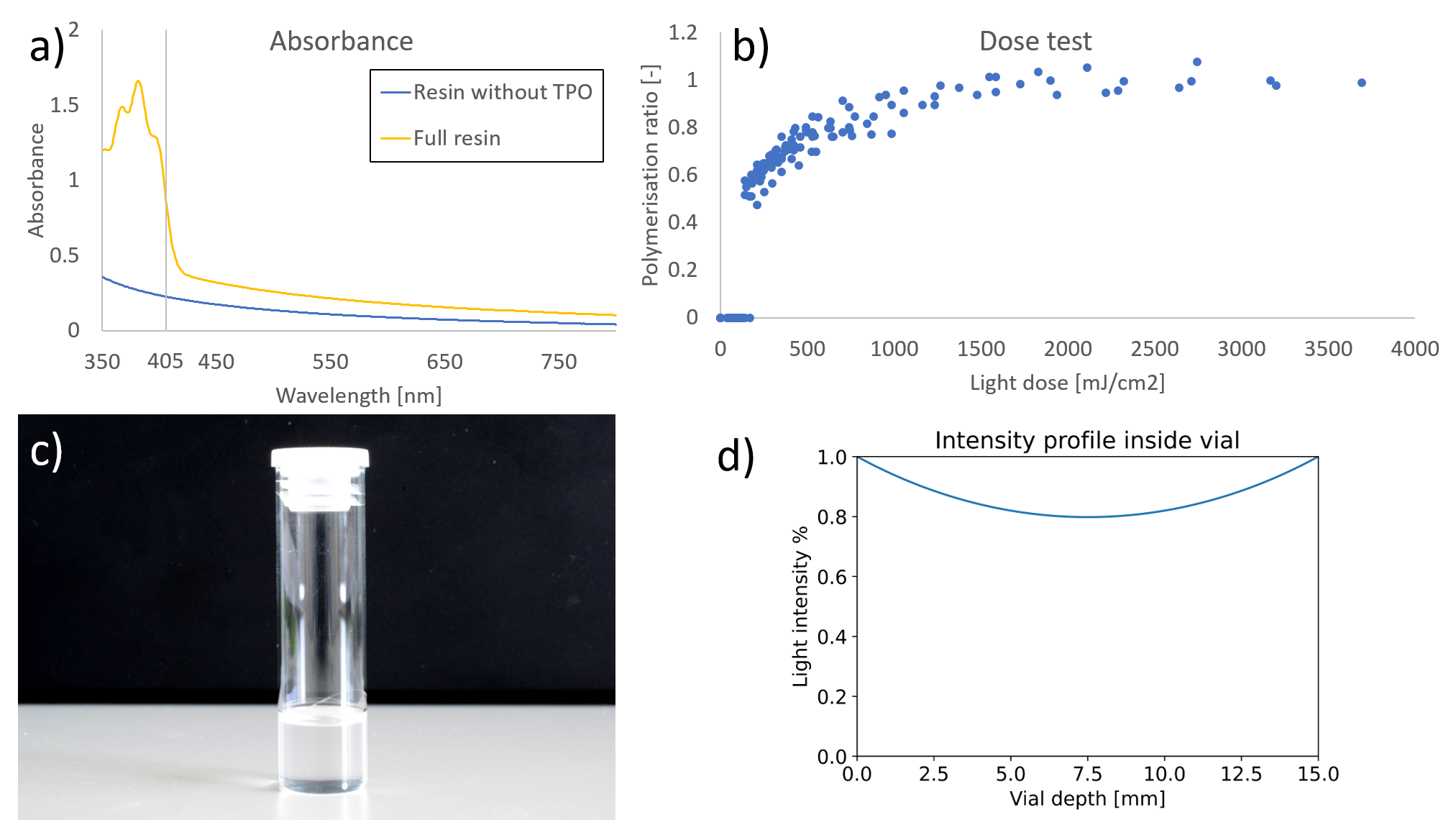}
    \caption{Preceramic resin characterisation.a) Absorbance measurements of the resin with and without the photoinitiator. (measured in 10mm wide vials). b) non-linear response of the resin to the light dose. c) Picture of a 16,5mm print vial with transparent resin. d) attenuation profile of the resin over the vial's width.}
    \label{resin}
\end{figure}

\paragraph{}The dose threshold is quantified by performing a dose test. The test consists in projecting round dots onto a 1mm thin vial filled with the resin. The dots have all a diameter of \(500\mu m\) and vary in intensity along one axis and vary in exposure time along the other (see figure \ref{resin} d)). This array of dots tests simultaneously different doses and this affects the size of the actually printed round dots. The polymerization ratio is determined as the size of the printed dot divided by the size of the projected dot. Some dots did not polymerize at all and some are bigger than the projection. This allows to determine the dose threshold of the resin (see figure \ref{resin} b)).

\subsection*{Rheology}

\begin{figure}[H]
    \centering
    \includegraphics[width=0.7\textwidth]{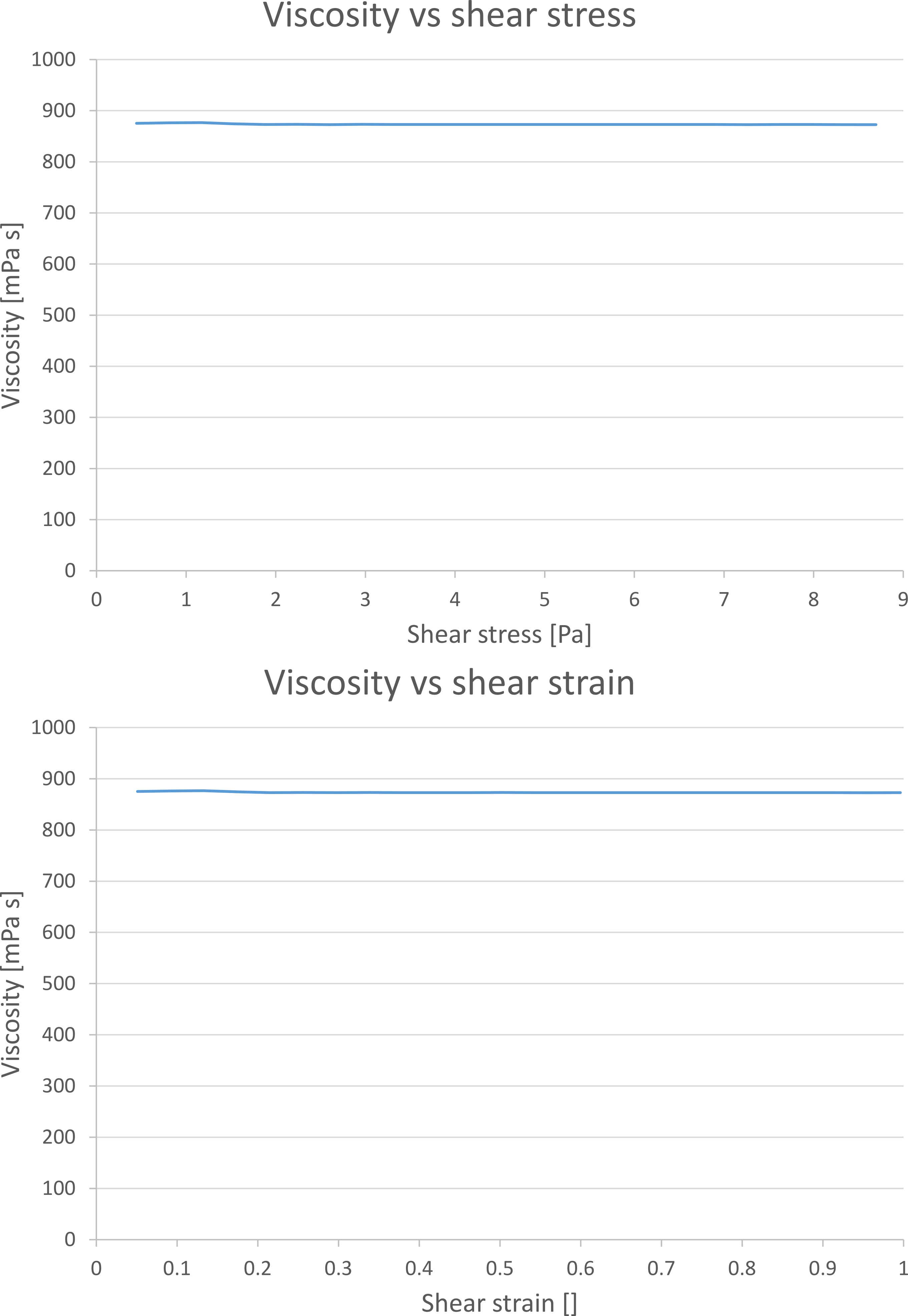}
    \caption{Plots of the rheology measurements. Top shows viscosity over shear stress. Bottom shows viscosity over shear strain}
    \label{viscosity}
\end{figure}

\subsection*{Shrinkage}

\begin{figure}[H]
    \centering
    \includegraphics[width=1\textwidth]{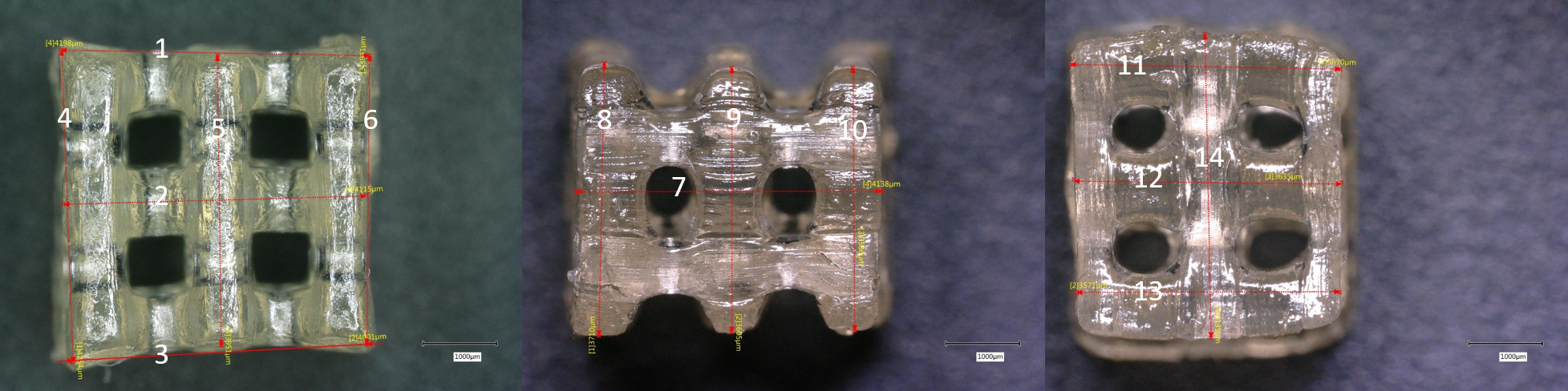}
    \caption{Images of the measurements taken for the shrinkage assessment.}
    \label{shrinkage}
\end{figure}

\paragraph{}To assess the shrinkage caused by the pyrolysis step, the square woodpiles were measured at different locations on all three main axes before and after the pyrolysis. The measurements were differentiated between the Z direction (vertical direction in the printer) and the x-y which are virtually interchangeable regarding the printing orientation. The measurements showed no favored direction of shrinkage.

\section*{pH measurements for chemical resistance}

\begin{figure}[H]
    \centering
    \includegraphics[width=0.7\textwidth]{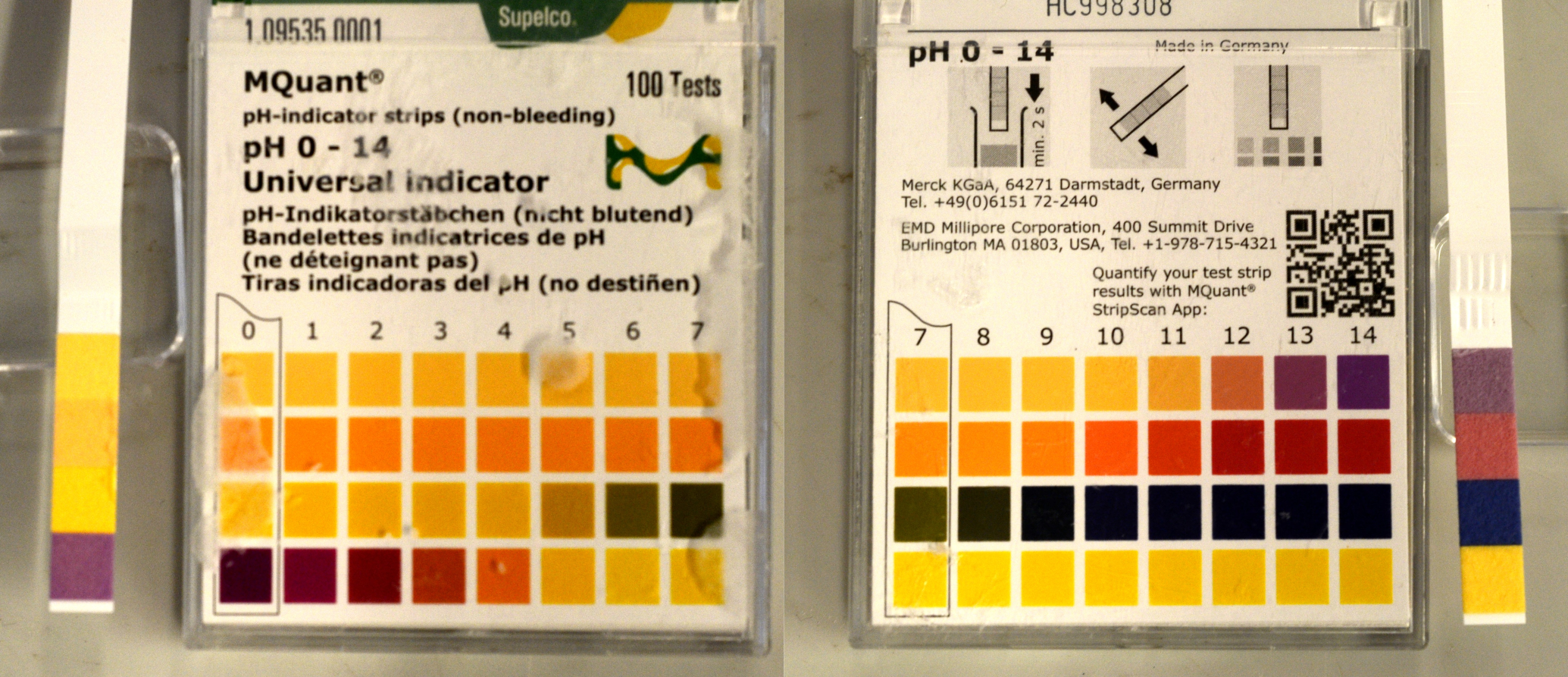}
    \caption{Photograph images of the pH measurement of the HCl bath (left) and KOH bath (right)}
    \label{pH}
\end{figure}

\section*{Micro computed tomography}

\begin{figure}[H]
    \centering
    \includegraphics[width=1\textwidth]{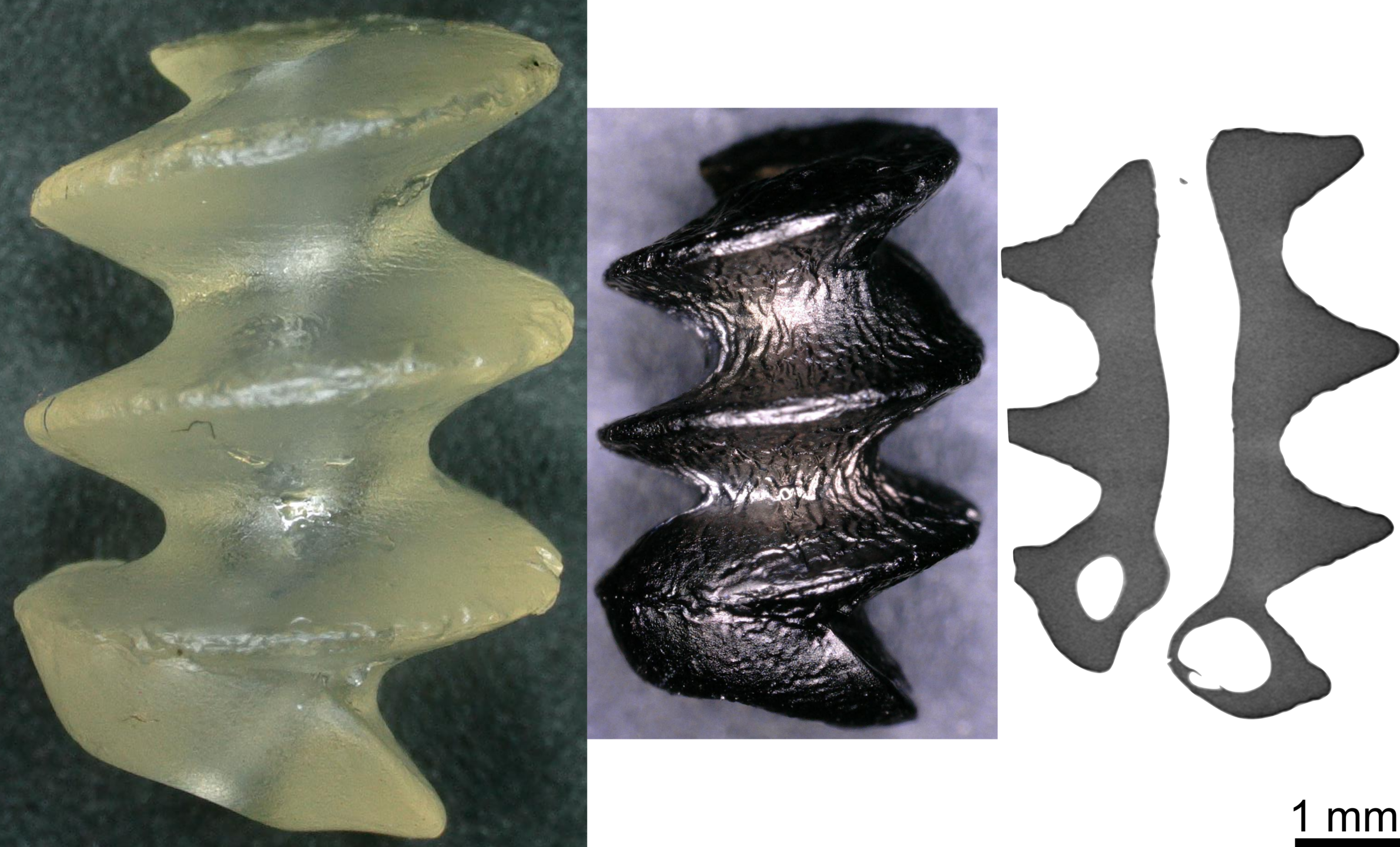}
    \caption{Micro computed tomography slice of a screw with a channel}
    \label{uCT}
\end{figure}

\section*{Pyrolysis}
\begin{figure}[H]
    \centering
    \includegraphics[width=1\textwidth]{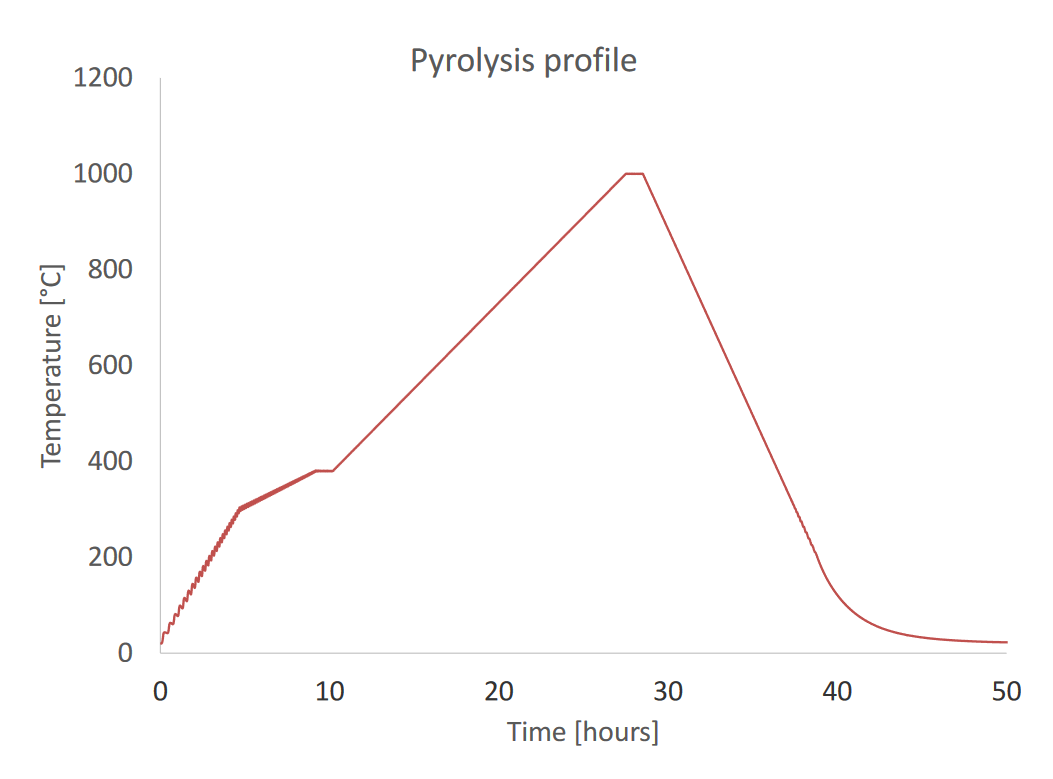}
    \caption{Measured pyrolysis temperature profile}
    \label{pyrolysis}
\end{figure}

\paragraph{}Figure \ref{pyrolysis} shows the temperature profile followed to transform the green bodies into polymer derived ceramics. The main dwell time is at $T=1000^{\circ}$C, which is the temperature at which the preceramic polymer loses all its organic components and becomes silicon oxycarbide. The crosslinker also has to be decomposed and evacuated, which is mostly done at the other dwell time at $T=380^{\circ}$C for one hour. Since the crosslinker is completely decomposed, this resin composition has a lot of matter to be outgassed. This makes the pyrolysis challenging and very prone to cracks and swelling. To hinder these negative effects, the heating and cooling rates are very slow to provide more time for the gases to escape. This makes the pyrolysis more gentle and allows for a better success rate. The first heating ramp is done at 1K/min because no component gets decomposed under $300^{\circ}$C. The second ramp is done at 0.3K/min. At this step, we are approaching the crosslinker decomposing temperature of $380^{\circ}$C and this has to be done gently. After the first dwell time of 1 hour, most of the crosslinker is evacuated and so the heating ramp can be brought up to 0.6K/min. After the second dwell time, the parts have become ceramics, so the cooling step can be faster. The cooling rate is 1.3K/min. It is visible in Figure \ref{pyrolysis} that between room temperature and $300^{\circ}$C, the temperature controller does not follow precisely the temperature profile which is set. This does not influence the result since at these temperatures, nothing of importance happens. After around 40 hours, the cooling is not fast enough anymore to follow the desired rate. Since the furnace does not have active cooling, the temperature decays exponentially.The decomposing temperature of the crosslinker was determined by thermogravimetric analysis.

\section*{Thermogravimetric Analysis}

\begin{figure}[H]
    \centering
    \includegraphics[width=1\textwidth]{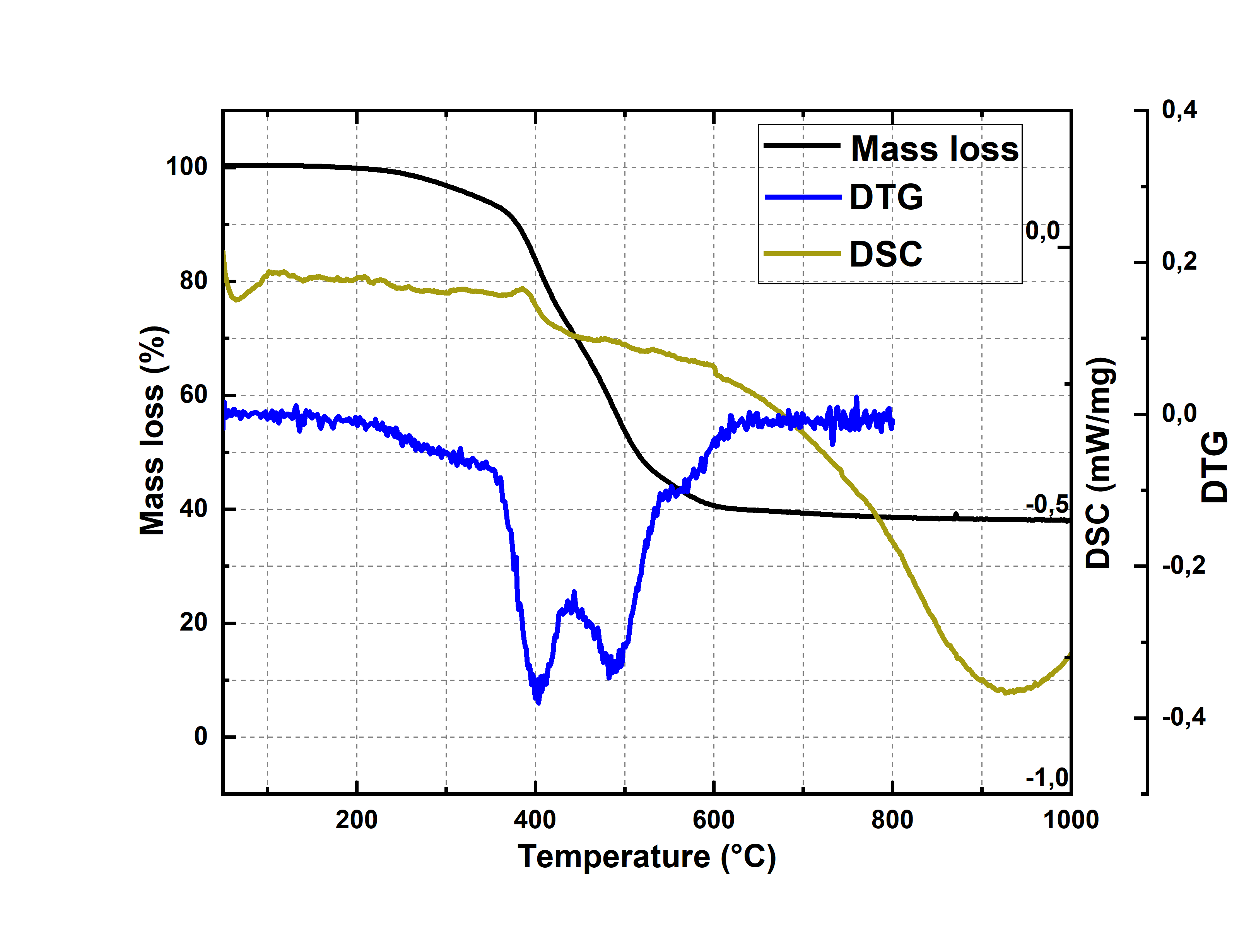}
    \caption{Mass loss in percentage (black line), Derivative thermogravimetry (DTG- blue line) and  Differential Scanning Calorimetry (DSC- gold line) profiles.}
    \label{TGA analysis}
\end{figure}

Figure \ref{TGA analysis} shows the results of Thermogravimetric Analysis. The samples were heated with a heating rate of 5 degrees/minute to maximum temperature under controlled argon atmosphere. During the measurement, the mass loss percentage and the Differential scanning Calorimetry are obtained. In the calculated Derivative thermogravimetry profile (DTG- blue line) two major decomposition intervals are observed with the appearance of two peaks, first one starting at 375$^{\circ}$C and the second one above 470$^{\circ}$C. The second one is common for PDCs and it appears to be completed above 600$^{\circ}$C. The first mass loss is mainly due to the decomposition of BDDA, which starts above 370$^{\circ}$C. This explains the dwelling step applied at that temperature (to allow smooth release of all the volatile gases) in the pyrolysis profile. Additionally, this step contributes to reduce significantly the formation of bubbles.Similarly, in DSC (gold line) we can observe a small peak around this temperature, which is attributed to this process. The mass loss is completed at  $\sim$ $600^{\circ}$C and there are no more mass losses as seen from the TG curve (mass loss \%). 

\end{document}